%% file: main.tex
\documentclass[twocolumn]{aastex62}
\pdfoutput=1 
\usepackage{amsmath,amstext}
\usepackage[figure,figure*]{hypcap}

\newcommand{\mesa}{\texttt{MESA}}
\newcommand{\gyre}{\texttt{GYRE}}
\newcommand{\iso}[2]{{}^{#2}\mathrm{#1}}
\newcommand{\pcap}[5]{\iso{#1}{#2} + p &\rightarrow{}&\iso{#3}{#4} + #5}
\newcommand{\pgamma}[4]{\pcap{#1}{#2}{#3}{#4}{\gamma}}
\newcommand{\palpha}[4]{\pcap{#1}{#2}{#3}{#4}{\alpha}}

\newcommand{\bdecay}[4]{\iso{#1}{#2} &\rightarrow{}&\iso{#3}{#4} + e^+ + \nu_e}
\newcommand{\sigv}[0]{\left\langle{}\sigma{} v\right\rangle{}}

\shorttitle{Non-Radial Pulsations in Post-Outburst Novae}
\shortauthors{Wolf et al.}

\received{November 20, 2017}
\accepted{February 2, 2018}
\submitjournal{ApJ}

\begin{document}

\title{Non-Radial Pulsations in Post-Outburst Novae}

\author[0000-0002-6828-0630]{William M. Wolf}
\affiliation{Department of Physics, University of California, Santa Barbara, CA 93106, USA}
\affiliation{School of Earth and Space Exploration, Arizona State University, Tempe, AZ 85287, USA}
\author[0000-0002-2522-8605]{Richard H.~D.~Townsend}
\affiliation{Department of Astronomy, University of Wisconsin-Madison, Madison, WI 53706, USA}
\author{Lars Bildsten}
\affiliation{Department of Physics, University of California, Santa Barbara, CA 93106, USA}
\affiliation{Kavli Institute for Theoretical Physics, University of California, Santa Barbara, CA 93106, USA}

\correspondingauthor{William M. Wolf}
\email{wmwolf@asu.edu}

\begin{abstract}

After an optical peak, a classical or recurrent nova settles into a brief (days
to years) period of quasi-stable thermonuclear burning in a compact
configuration nearly at the white dwarf (WD) radius. During this time, the
underlying WD becomes visible as a strong emitter of supersoft X-rays.
Observations during this phase have revealed oscillations in the X-ray emission
with periods on the order of tens of seconds. A proposed explanation for the
source of these oscillations are internal gravity waves excited by nuclear
reactions at the base of the hydrogen-burning layer. In this work, we present
the first models exhibiting unstable surface $g$-modes with periods similar to
oscillation periods found in galactic novae. However, when comparing mode
periods of our models to the observed oscillations of several novae, we find
that the modes which are excited have periods shorter than that observed.

\end{abstract}

\input{introduction}
\input{nova_calculations}
\input{pulsational_analysis}
\input{nova_modes}
\input{discussion}
\input{conclusions}

\acknowledgments
This work was supported by the National Science Foundation under grants PHY
11-25915, ACI 13-39581, ACI 13-39606, and ACI 16-63688, as
well as by NASA under TCAN grant numbers NNX14AB53G and NNX14AB55G. We thank 
Steve Kawaler for helpful discussions regarding his previous work on 
planetary nebula nuclei. We also thank the anonymous referee for helpful 
comments and questions that improved the manuscript.
\software{\mesa{} \citep{2011ApJS..192....3P, 2013ApJS..208....4P, 
2015ApJS..220...15P}, \gyre{} \citep{2013MNRAS.435.3406T,Townsend2017}, Numpy
\citep{doi:10.1109/MCSE.2011.37}, Matplotlib \citep{doi:10.1109/MCSE.2007.55},
and IPython \citep{doi:10.1109/MCSE.2007.53}.}

\bibliographystyle{aasjournal}
\bibliography{bibliography}

\appendix

\input{appendix_phase}
\listofchanges
\end{document}

%% file: introduction.tex
\section{Introduction} 
\label{sec:introduction}

A nova is an optical event caused by a thermonuclear runaway on the surface of
a white dwarf (WD) \citep{1978ARA&A..16..171G}. The thermonuclear runaway
drives a rapid expansion of the WD where it shines brightly in the optical and
loses much of its hydrogen-rich envelope via some combination of dynamical
ejection, optically-thick winds, and/or binary interactions. Eventually enough
mass is lost from the envelope so that the photospheric luminosity matches the
nuclear burning luminosity and the WD radius recedes to a more compact
configuration \citep{2014ApJ...793..136K}. Hydrogen burning does not cease,
though, as a remnant envelope is slowly burned over days to decades. The hot
and compact WD shines brightly in the UV and soft X-rays, appearing very
similar to a persistent supersoft source (SSS) \citep{2013ApJ...777..136W}.
Dozens of SSSs from post-outburst novae are seen in M31
\citep{2010A&A...523A..89H, 2011A&A...533A..52H, 2014A&A...563A...2H,
2006ApJ...643..844O, 2010ApJ...717..739O} and the Milky Way \citep[and
references therein]{2011ApJS..197...31S} every year.

Many, if not all, SSSs exhibit periodic oscillations in their X-ray light curve
with periods ($P_\mathrm{osc}$) in the range of 10-100 seconds, whose precise
origin is unclear \citep[and references therein]{2015A&A...578A..39N}.
\cite{2014MNRAS.437.2948O} argue that in the case of Cal 83, its 67 s period
could be the rotational period of the WD. \cite{2015A&A...578A..39N} point out
that the observed drift of the precise $P_\mathrm{osc}$ of $\pm 3$ s  can't be
easily explained by accretion spin-up or spin-down (due to high inertia of  the
WD) or by Doppler shifts of the emitting plasma due to the orbital motion.
Furthermore, the $P_\mathrm{osc}=67$ s  of Cal 83 is the longest in the known
sample, so  other WDs would need to be rotating even more rapidly. While the
rotation rates of accreting WDs are still not well understood, spectroscopic
measurements to date do not point to rapid rotation
\citep{1999PASP..111..532S,2012ApJ...753..158S,2016MNRAS.457.1828K}.

Rotation is thus not a very promising mechanism for explaining these
oscillations, though it cannot be ruled out until an independent determination
of the WD rotation period is obtained in an oscillating SSS. A more promising
explanation first proposed by \cite{2003ApJ...584..448D} is that the
oscillations are caused by non-radial surface $g$-modes excited by the
$\epsilon$-mechanism \added{(driving due to compressional sensitivity of the nuclear burning rate)} at the base of the hydrogen burning layer. However, the
oscillations observed by \cite{2003ApJ...584..448D} for nova V1494 Aquilae were
much longer. At $P_{\mathrm{osc}}\approx 2500$ s, these modes were more
credibly explained as being driven by the $\kappa$-mechanism \added{(driving due to compressional sensitivity of the opacity)}, where an
ionization zone, rather than  temperature-sensitive burning, is the source of
an instability. Indeed, longer periods ($\sim10-100$ minutes) have been
observed in Cal 83 \citep{1987ApJ...321..745C,2006AJ....131..600S} and nova
V4743 Sgr \citep{2003ApJ...594L.127N}, all consistent with oscilations most
similar to \replaced{ZZ Ceti, driven by the outermost convection zone}{GW Vir, driven by the ionized carbon and oxygen}. These 
longer-period oscillations are not the focus of this work.

The expected $P_\mathrm{osc}$ for $\epsilon$-mechanism-driven $g$-modes was
estimated in \cite{2015A&A...578A..39N} for a typical WD mass, envelope mass,
and a constant-flux radiative envelope to be on the order of 10 s, in great
agreement with the observed periods. Their calculation, however, could not
assess whether the mode would grow unstably or damp out.

The configuration of a thin hydrogen-burning radiative envelope on a WD is
similar to early planetary nebulae nuclei, as explored by
\cite{1988ApJ...334..220K}. With a detailed non-adiabatic pulsational
analysis, \cite{1988ApJ...334..220K} found that $g$-modes were indeed excited
by the $\epsilon$-mechanism. In a 0.618 $M_\odot$ planetary nebula nucleus
model, higher-order modes with $P_\mathrm{osc} \approx200\ \mathrm{s}$ were
excited first when the luminosity was around $\log L/L_\odot\approx 3.1$, and
lower order modes with $P_\mathrm{osc} \approx 70\ \mathrm{s}$ only being
excited after the luminosity dropped to $\log L/L_\odot \approx 2.6$.

Encouraged by the promising results of \cite{1988ApJ...334..220K} and
\cite{2015A&A...578A..39N}, we present in this paper the first detailed 
non-adiabatic calculations of the unstable modes in post-outburst nova models
using the open source stellar evolution code \mesa{} \texttt{star} \citep[rev.\
9575;][]{2011ApJS..192....3P, 2013ApJS..208....4P, 2015ApJS..220...15P} and the
accompanying non-adiabatic stellar pulsation tool \gyre{} 
\citep{2013MNRAS.435.3406T,Townsend2017}.  In \S \ref{sec:stellar_models} we
explain the simulation details to obtain post-outburst nova models from \mesa{}
\texttt{star} for input into \gyre. Then in \S \ref{sec:pulsational_analysis}
we discuss mode propagation in our models and compare to previous simulations
of oscillations in a planetary nebula nucleus. In \S
\ref{sec:supersoft_nova_modes}, we present the periods and growth timescales of
the modes calculated by \gyre{} from the nova models. We comment on how these
modes compare to observed oscillation periods in \S \ref{sec:discussion} before
summarizing in \S \ref{sec:conclusions}.

%% file: nova_calculations.tex
\section{Stellar Models} 
\label{sec:stellar_models}

To generate models for use in pulsational analysis, we use the  \mesa{}
\texttt{star} code. Specifically, we use an inlist based on the \texttt{nova}
test case scenario, which in turn was based off of the nova calculations of
\cite{2013ApJ...777..136W}. In these models, hydrogen-rich material is accreted
at a rate of $10^{-9}~M_\odot~\mathrm{yr^{-1}}$, which is a typical rate
expected for cataclysmic variables \citep{2005ApJ...628..395T}. Mass loss was
handled by the built-in super-Eddington wind scheme described in
\cite{2013ApJ...762....8D} and \citet{2013ApJ...777..136W}, as well as a
modified version of the built-in Roche lobe overflow mass loss scheme.

\begin{figure*}[t!]
  \plottwo{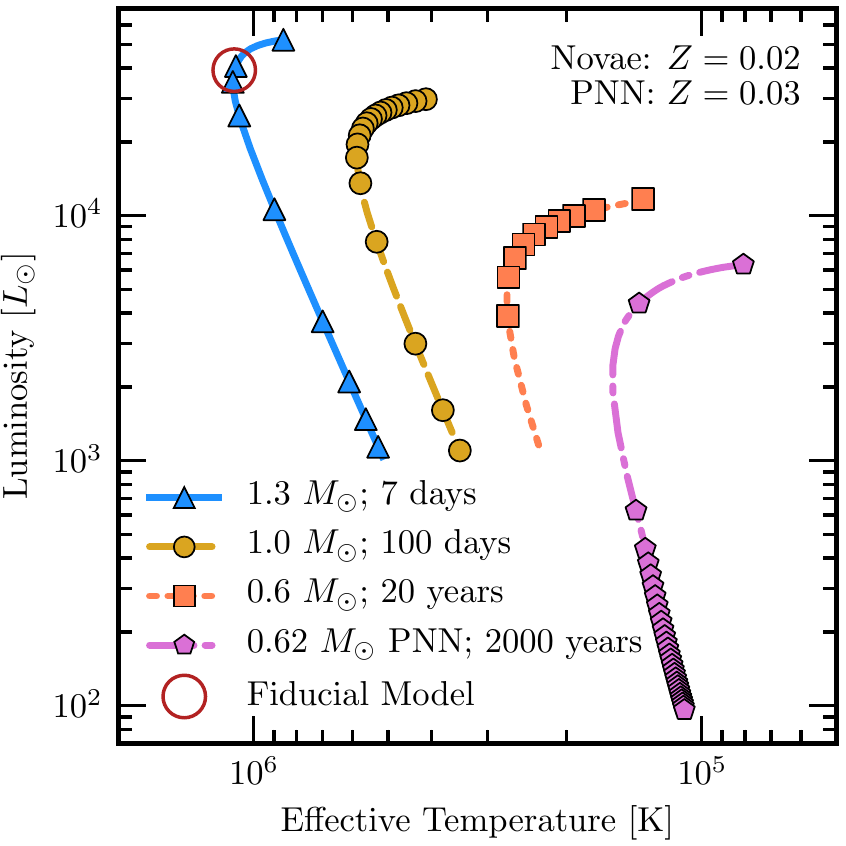}{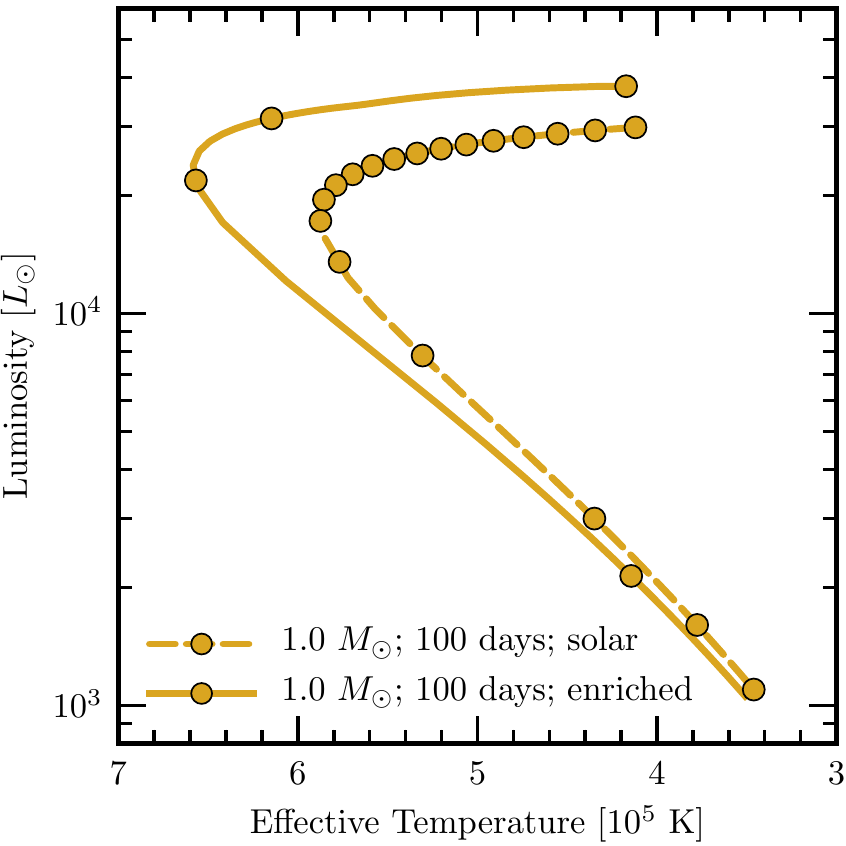}
  \caption{Evolution of all stellar models through the HR diagram. Different
  markers separate equal times of evolution. For example, between two yellow
  circles, 100 days have elapsed. \textbf{Left:} The three nova models that
  accrete solar composition material from the end of mass loss until their
  luminosities reach $10^3~L_\odot$. Also shown is the $M=0.6172~M_\odot$ 
  planetary nebula nucleus introduced in Section 3. The maroon circle indicates
  a fiducial model of the 1.3 $M_\odot$ nova that we use as an example later in
  the paper.
  \textbf{Right:} Comparison between the $1.0~M_\odot$ nova models accreting
  solar composition and 25\% core composition, 75\% solar composition material.
  Again, markers along each track mark intervals of equal time.}
  \label{fig:hr}
\end{figure*}

The precise nature of the mass loss is not important because mass is lost in
some form until the hydrogen rich layer is reduced to the maximum mass that can
sustain steady hydrogen burning in a compact form, which is a function
primarily of the WD mass. At this point the WD shrinks and enters its
post-outburst phase, as found by  \cite{2013ApJ...777..136W,
2014ApJ...793..136K}. The precise nature of the mass loss greatly affects
properties of the nova at the time of optical peak, which we are not interested
in. However, extra mass loss in excess of that required to reduce the hydrogen
layer mass down to a stable burning mass can truncate the duration of the 
post-outburst phase. To create the most favorable conditions for mode
excitation, we shut off mass loss or gain once the WD shrinks to radii similar
to the reddest steady-state burners found by \cite{2013ApJ...777..136W}. In
general, super Eddington winds dominate mass loss for novae on higher-mass WDs,
and Roche lobe overflow dominates mass loss for novae on the lowest-mass WD.

These models are non-rotating,
though rotationally-induced instabilities can be responsible for mixing between
core and accreted material \citep{1983ApJ...273..289M, 1987ApJ...318..316L,
1987ApJ...321..394S}. Rotation may also affect the stability and structure of
$g$-modes in a
stellar model, so we discuss the effects of modest rotation on the expected
modes in \S \ref{sec:supersoft_nova_modes}. No diffusion is allowed,
though at this high of an accretion rate, its effects on metal enrichment of
the thermonuclear runaway would not be very pronounced
\citep{1992ApJ...388..521I,1995ApJ...445..789P,2005ApJ...623..398Y}. Finally,
we do not allow for any \replaced{convective undershoot}{turbulent mixing at convective boundaries 
(i.e. undershoot/overshoot)} during the thermonuclear runaway,
which would also act to enhance the ejecta with metals
\citep{2010A&A...513L...5C,  2011A&A...527A...5C, 2011Natur.478..490C,
2012MNRAS.427.2411G}. Mixing due to rotational instabilities, diffusion, and/or
convective \replaced{convective dredge-up}{boundary mixing} are all causes of the metal enhancement of nova ejecta
indicated by optical and UV spectra \citep{1998PASP..110....3G,
2013ApJ...762..105D} as well as evidence for dust formation
\citep{1970ApJ...161L.101G, 1978ApJ...219L.111N, 1980ApJ...237..855G}.

Rather than considering how exactly to parameterize and combine the mixing
effects of rotational, diffusion-induced, and turbulent instabilities, we
instead include a model where the accreted material is 25 percent core
material, where ``core composition'' is defined as the composition sampled
where the helium mass fraction first drops below one percent. The remaining 75
percent of accreted material is solar composition.

All inlists, models, and additional code used to produce these models will be
posted on the \mesa{} users' repository,\ \url{mesastar.org}.

In total, four models were calculated: pure solar material accretion models for
WD masses of 0.6 $M_\odot$, 1.0 $M_\odot$, and 1.3 $M_\odot$ and a 
metal-enriched accretion model for a 1.0 $M_\odot$ WD. The starting models 
were the endpoints of the similar nova simulations carried out by
\cite{2013ApJ...777..136W}. The solar composition models were  evolved through
2-3 nova cycles to erase initial conditions, while the metal-rich models were
evolved through several flashes at an intermediate metallicity before being
exposed to 25\% enrichment to ease the transition. In all cases, model
snapshots at every timestep after the end of mass loss to the end of the SSS
phase were saved and form the basis for the analysis in the rest of this work.

Figure \ref{fig:hr} shows the evolution of these nova models as well as a
planetary nebula nucleus model with $M=0.617~M_\odot$ introduced in \S
\ref{sec:pulsational_analysis} through the HR diagram. The general trends are
that higher mass WDs and more metal-rich accretion give faster, bluer, and more
luminous evolution. Note that the markers break the evolution into stretches of
equal duration, but the actual timesteps taken in the evolution were much
shorter, taking somewhere between 30 and 60 timesteps to get through the SSS
phase. Also indicated in Figure~\ref{fig:hr} is the location of a fiducial
model from the $1.3~M_\odot$ simulation. We will refer to this model in
subsequent sections as an example case for mode analysis.

%% file: pulsational_analysis.tex
\section{Non-Radial Pulsation Analysis} 
\label{sec:pulsational_analysis}

With model snapshots of each of the novae throughout the SSS phase, we can use
\gyre{} to determine their oscillation modes, focusing only on the $\ell=1$
(dipole) modes. We begin by looking at the adiabatic properties of our
fiducial model before delving into non-adiabatic analyses.

\subsection{Adiabatic Pulsation} 
\label{sub:adiabatic_pulstaion}
\gyre{} analyzes a stellar model to find its radial and non-radial
pulsation modes. While a non-adiabatic calculation is required to determine
which of these modes are excited in a given stellar model, we can
learn a lot from simpler adiabatic calculations to see what modes are 
available for excitation.

\begin{figure}
  \includegraphics[width=\columnwidth]{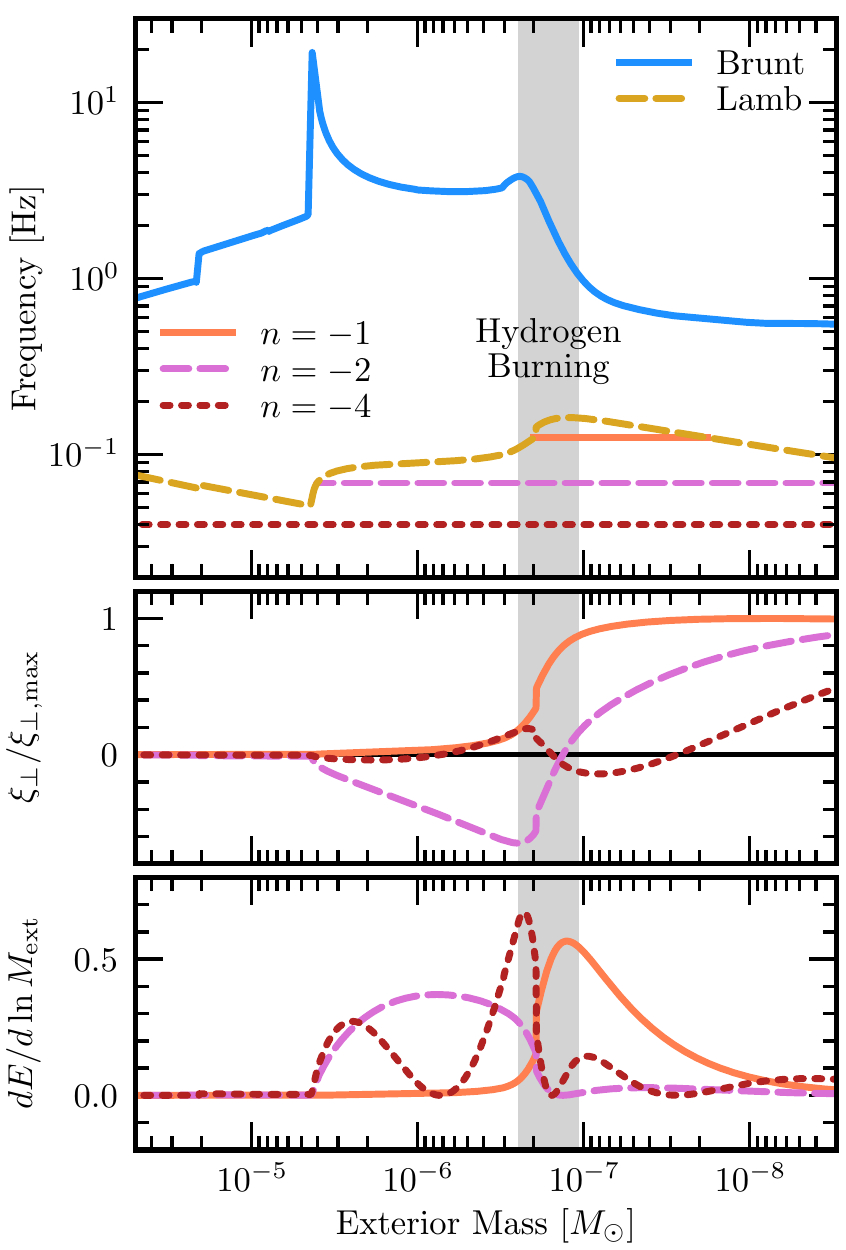}
  \caption{Profiles of the fiducial 1.3 $M_\odot$ model introduced in 
  Figure~\ref{fig:hr}.
  \textbf{Top panel}: Propagation diagram for our fiducial 1.3 $M_\odot$ 
  post-outburst nova model in its outermost $10^{-4}~M_\odot$. The shaded 
  region indicates the region over which 80\% of the stellar luminosity is
  generated by CNO burning. Regions where the $n=-1, -2$ and $-4$
  modes can propagate are plotted as horizontal lines at their respective
  frequencies.
  \textbf{Middle panel}: Eigenfunctions for the same three modes. Horizontal
  displacement dominates over radial displacement for these modes, so only
  the horizontal displacement is shown, normalized to a maximum of unity.
  \textbf{Bottom panel}: Mode inertia of these same modes expressed as $dE/d\ln M_{\mathrm{ext}}$, the derivative of the inertia with respect to 
  $\ln [M(R) - M(r)] = \ln M_{\mathrm{ext}}$ so that equal areas under the
  curve indicate equal mode inertias. This is again normalized to integrate to unity.}
  \label{fig:prop}
\end{figure}

We aim to explain the observed oscillations as $g$-modes in the outer
atmosphere, so some $g$-modes in our model must ``live'' in the outermost 
parts of our model. The upper panel of Figure~\ref{fig:prop} shows a
propagation diagram of our fiducial 1.3 $M_\odot$ model during its SSS phase.
Also indicated is the region of strong hydrogen burning, where we expect mode
driving to occur.

After using \gyre{} to search for the eigenmodes of this model, we indeed find
$g$-modes that live in the outer atmosphere with periods on the order of a few
to tens of seconds. Horizontal displacement eigenfunctions for the $g$-modes
with radial orders $n=-1, -2$, and $-4$ (in the Eckart-Osaki-Scuflaire
classification scheme, as modified by \citet{2006PASJ...58..893T}) are shown
in the middle panel of Figure~\ref{fig:prop}. The frequencies of these modes
are also shown as horizontal lines spanning their allowed propagation regions
(where their frequencies lie below both the Lamb and Brunt-V\"ais\"al\"a
frequencies) in  the upper panel.  The bottom panel shows the distribution of
inertia in these modes (normalized to integrate to unity), confirming 
that the modes indeed exist only within their allowed propagation regions. We
see that the lowest order mode lives mostly in the burning region and the
lower-density region above it. This makes this mode comparatively easier to
excite than the other two, which have much of their energy in the 
higher-density helium-rich region below.

These are merely the modes in which the star is able to pulsate. To excite 
one, a driving force must do work on the mode, and a non-adiabatic caluclation
is required to find such unstable modes. We discuss the relevant driving force
and our non-adiabatic calculations next.

\subsection{Non-adiabatic Pulsations and the $\epsilon$-Mechanism} 
\label{sub:non_adiabatic_pulsations_and_the_}
The driving force relevant to novae in the SSS phase as well as planetary
nebula nuclei is the $\epsilon$-mechanism. In the $\epsilon$-mechanism, the
nuclear energy generation rate per unit mass $\epsilon$ is enhanced during a
compression and attenuated during rarefaction. In this way, heat is added
near the maximum temperatre of the cycle and removed near the minimum
temperature, creating a heat engine that converts thermal energy into work
\citep{1926ics..book.....E}.

This phenomenon requires temperature sensitivity to produce feedback between
the pulsation and $\epsilon$. For temperatures of interest to this work
($T\lesssim 10^8$ K), the CNO cycle is not yet beta-limited, and we still have
$\epsilon\propto T^{9-14}$, so the $\epsilon$-mechanism can still be relevant.

There is, however, a minor complication. With periods on the order of tens of
seconds, oscillations in temperature and density occur on the same timescales
as the lifetimes of isotopes in the CNO cycle \citep{1988ApJ...334..220K}.
This leads to lags between the phases of maximum
temperature/density and the phase of maximum energy generation. As a
result, the temperature and density sensitivities of the nuclear energy
generation rate will differ from those in a non-oscillating system at the same
average temperature and pressure.

The method for computing corrected partial derivatives of the energy
generation rate were presented in \cite{1988ApJ...334..220K}, but since that
work examined oscillations in a planetary nebula nucleus, which burns at a
lower temperature than our nova models, an assumption in that work does not
apply here. The details of how we calculate the partial derivatives and 
include them in \gyre{} are in Appendix \ref{sec:appendix_phase_lags}.

\begin{figure}
  \includegraphics[width=\columnwidth]{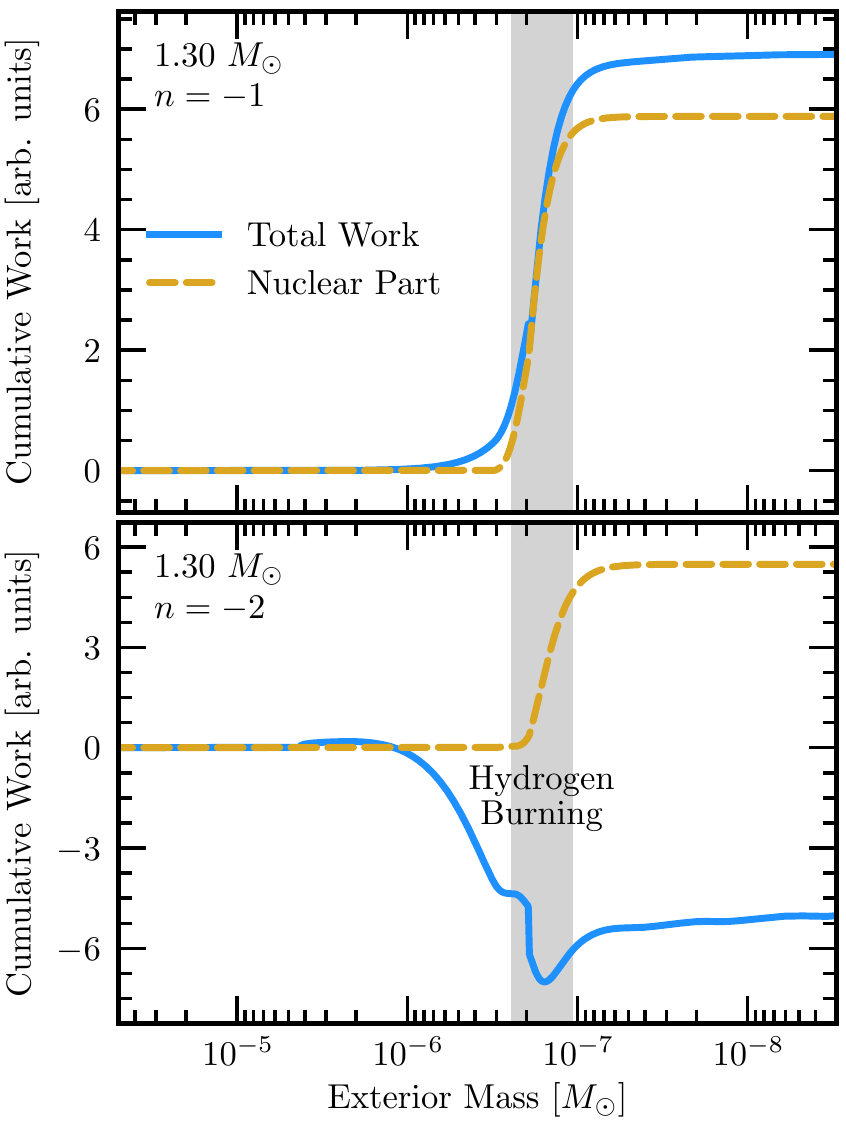}
  \caption{The cumulative integrated work done on the $n=-1$ (top panel) and
  $n=-2$ (bottom panel) modes in the fiducial model in arbitrary units as a
  function of the exterior mass $\Delta M_{\mathrm{ext}}(r)=m(R) - m(r)$. The solid blue line is the
  result of a fully nonadiabatic calculation, with the broken gold line being
  the contribution from the $\epsilon$-mechanism. The net positive work done in
  the top panel indicates that the $n=-1$ mode is unstable, while the net
  negative work in the bottom panel indicates that the $n=-2$ mode is stable
  despite the destabilizing (positive) contribution of the $\epsilon$ term.}
  \label{fig:work}
\end{figure}

A mode is excited when a driving mechanism does enough work on the mode to 
exceed the energy lost through damping mechanisms over one oscillation cycle 
\citet[chapter V]{1989nos..book.....U}. In Figure~\ref{fig:work}, we show the
cumulative work done on the $n=-1$ and $n=-2$ modes in our fiducial model. 
We show both the total cumulative work and only the work done by the 
$\epsilon$-mechanism. A net positive work indicates global mode driving and a net
negative work indicates global mode damping. Note that in both cases, the contribution from the $\epsilon$-mechanism is positive, so it is always a
driving force. However, in the $n=-2$ mode, nuclear driving is not strong 
enough to overcome other damping forces and the mode is globally damped. In the $n=-1$
mode, though, driving forces win and the mode is excited.

Notably, the total work done on the $n=-1$ mode exceeds that done by nuclear
driving alone, which means another mechanism is also contributing to the
instability. This mechanism is related to the steep luminosity gradient
present in the burning region (i.e.\ \emph{not} the $\kappa$-mechanism). We 
defer more exploration of this mechanism to subsequent work.

Before looking further at the modes excited in the nova models, we first 
analyze a planetary nebula nucleus model similar to that of
\citet{1988ApJ...334..220K} to verify that we obtain a similar set of excited
modes.

\subsection{Planetary Nebula Nucleus} 
\label{sub:planetary_nebula_nucleus}

The planetary nebula nucleus (PNN) model from \citet{1988ApJ...334..220K} was
created by first evolving a star with a ZAMS mass of a 3.0 $M_\odot$ star with
a metallicity of $Z=0.03$ to the AGB and then stripping its envelope gradually
away.

The \mesa{} test suite includes a test case, \texttt{make\_co\_wd},
which evolves a star to the AGB and through one thermal pulse from the helium
burning shell, and then greatly increases the efficiency of AGB winds to reveal
the WD. We used this test case as a basis and changed three controls to create
our PNN model. First, we set the metallicity to 0.03 instead of the test case's
default value of 0.02. Secondly, we evolve the model from the pre-main
sequence (rather than interpolating from a default suite of models) due to
the specific metallicity. Finally, we adusted the initial mass to
$3.30~M_\odot$ so that the final mass of $M = 0.6172~M_\odot$) closely 
resembled the mass of the PNN in \citet{1988ApJ...334..220K} of 
$M = 0.6185~M_\odot$.

Once the model reached an effective temperature greater than 10,000 K, we
changed its nuclear network to match the network used in the nova simulations
(\texttt{cno\_extras.net}). At $T_{\mathrm{eff}}=60,000$ K, we halted the
enhanced mass loss that accelerated the thermal pulse phase in order to resume
normal PNN evolution. We then saved profiles for pulsational analysis at every
timestep once the effective temperature exceeded 80,000 K, and we halted
evolution when the luminosity dropped below 100 $L_\odot$.

\begin{figure}
  \includegraphics[width=\columnwidth]{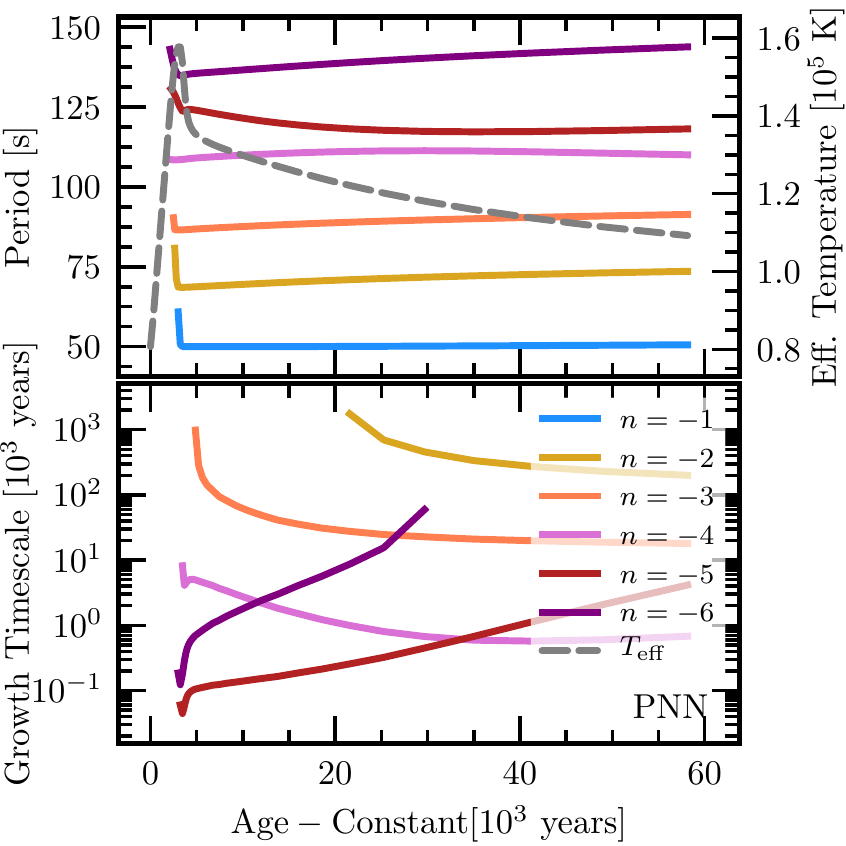}
  \caption{
  Evolution of $\ell=1$ $g$-modes in the planetary nebula nucleus model through
  the depletion of its hydrogen envelope. The top panel shows how the periods
  of the six lowest-order $g$-modes change in time. The effective temperature
  is also shown for comparison to evolution in the HR diagram. The bottom panel
  shows the evolution of the growth timescale for each mode if it is unstable.}
  \label{fig:pnn}
\end{figure}

The evolution of the model's $g$-mode properties through its PNN phase is shown
in Figure~\ref{fig:pnn} for six lowest-order modes. The first mode to be
excited was a $g$-mode with radial order $n=-6$. The period of this mode stayed
consistently near 150 s and its growth time stayed in the range of hundreds to
thousands of years (still shorter than the hydrogen-burning lifetime of the
PNN). The period agrees well with the $k=6$ column of Table 3 in
\citet{1988ApJ...334..220K}, but we find growth timescales that are longer by
one or more orders of magnitude with the mode being stabilized sooner than in
\citet{1988ApJ...334..220K}.

Other modes have matching or very nearly matching periods, but the growth times
we find are typically much longer than those of \citet{1988ApJ...334..220K}. In
addition to the modes shown in Figure~\ref{fig:pnn}, we see the $n=-7$ and
$n=-8$ modes excited, but not the $n=-9$ mode as in 
\citet{1988ApJ...334..220K}, consistent with the general trend of higher 
stability in our models.

We searched for modes both while accounting for the phase lags in the energy
generation rate and while not accounting for them. In both PNN and nova
models, adding in the effects of phase lags increases
growth times and stabilizes modes that would otherwise be unstable. This is
because the phase of peak heat injection is moved away from the phase of peak
temperature/density, weakening the heat engine set up by the 
$\epsilon$-mechanism.


%% file: nova_modes.tex

\begin{figure*}
  \plottwo{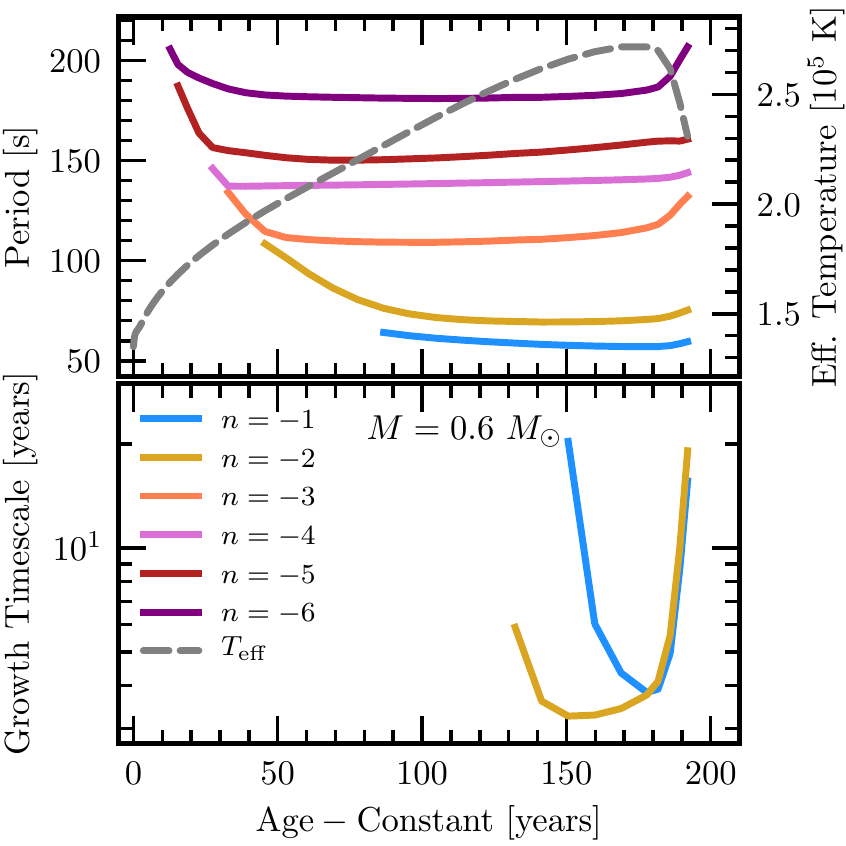}{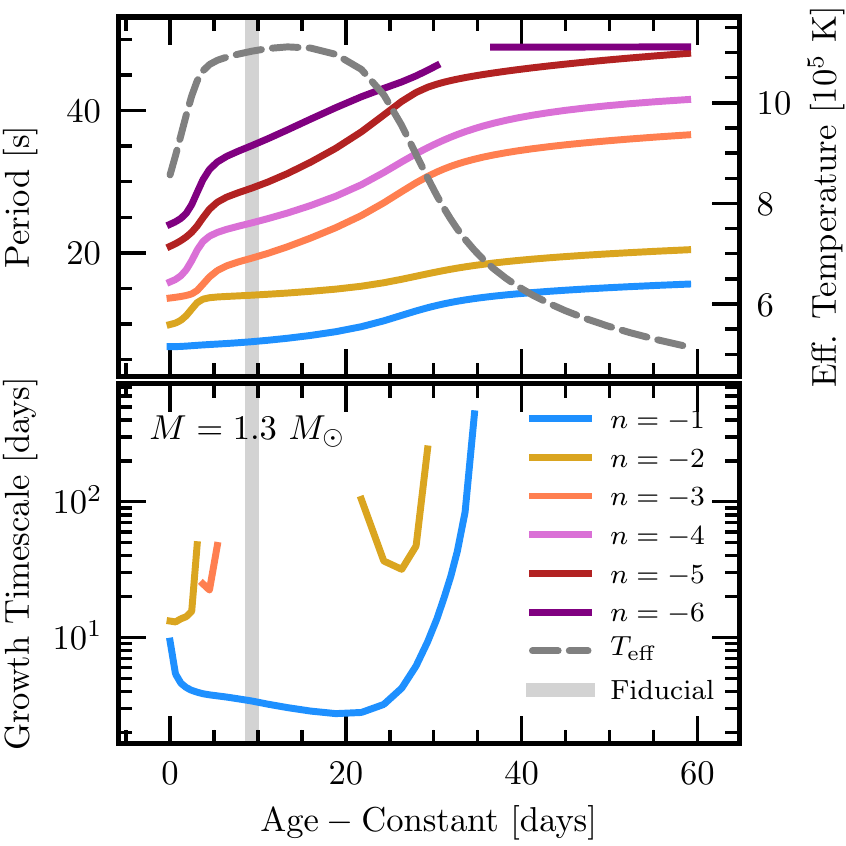}\\
  \vspace{0.2in}
  \plottwo{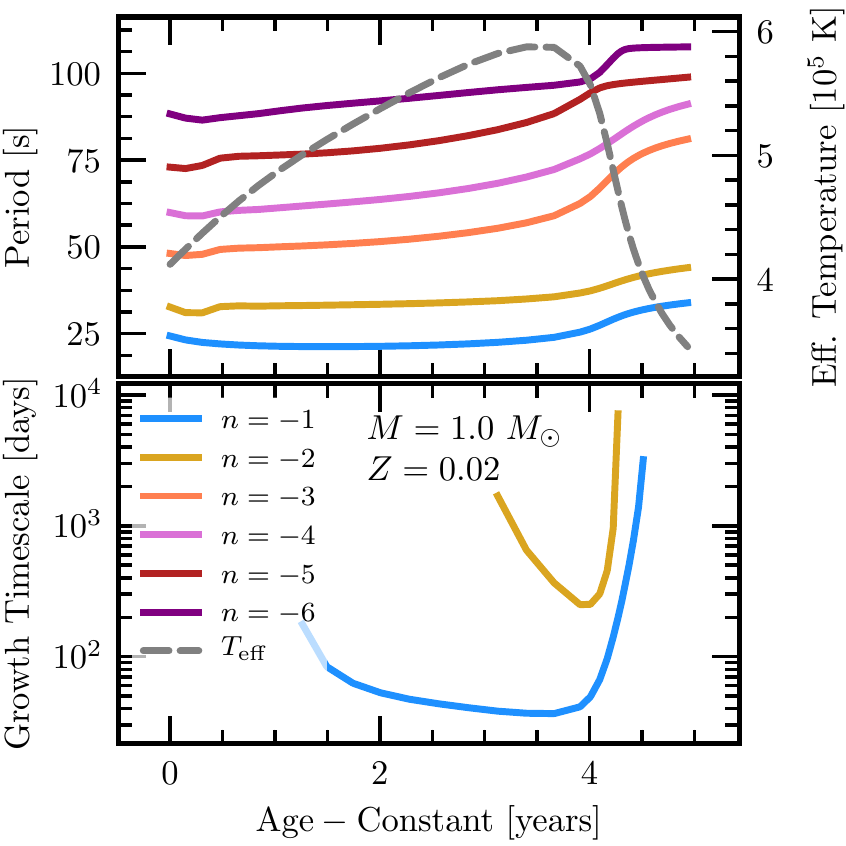}{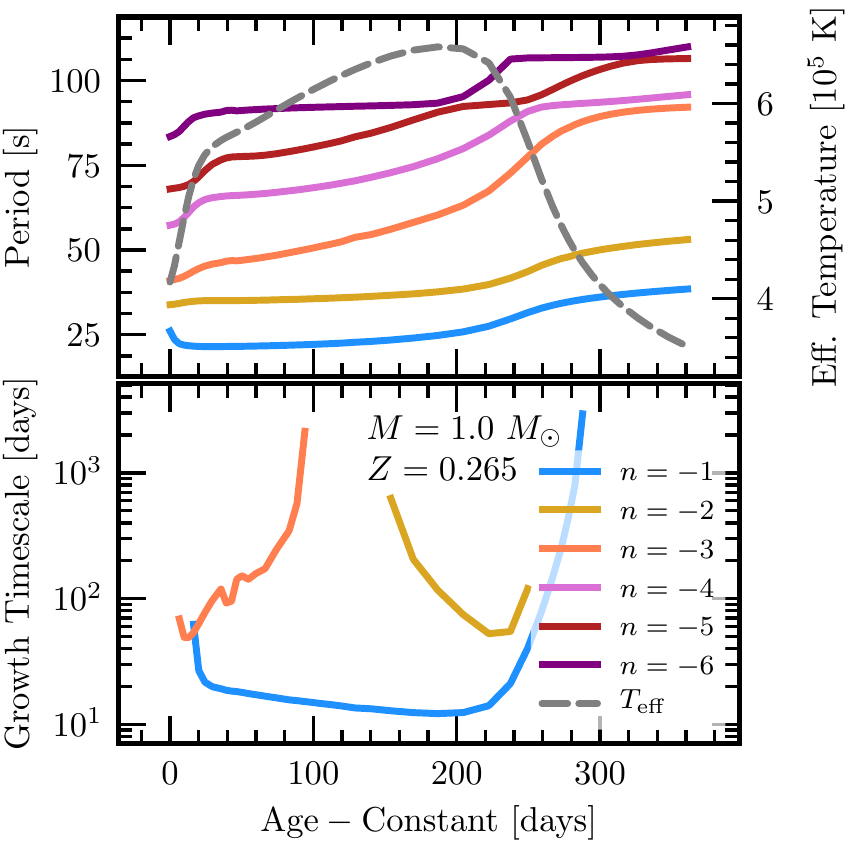}
  \caption{
  Evolution of the $\ell=1$ $g$-modes modes of each post-outburst nova model
  (masses and compositions indicated in each plot). Similar to
  Figure~\ref{fig:pnn}, the top panels show mode periods and effective
  temperatures while the bottom panels show growth timescales. Points in a top
  panel represent unstable modes only if an accompanying point at the same age
  and mode order appears in the lower panel. A gray vertical band in the
  $1.3~M_\odot$ plot indicates from where the fiducial model referenced
  elsewhere in this work is taken.}
    \label{fig:taus1}
\end{figure*}

\section{Supersoft Nova Modes} 
\label{sec:supersoft_nova_modes}
Figure \ref{fig:taus1} shows the evolution of the periods
of low-order $g$-modes in the post-outburst nova models as well as the
evolution of these modes' growth timescales. The effective temperature
evolution is also shown in these figures, revealing that the most rapid 
excitation occurs in the approach to the peak effective temperature at the 
"knee" of the HR diagram shown in Figure~\ref{fig:hr}.

We find unstable modes excited on timescales shorter than the supersoft phase
lifetime in all four nova models. Excited modes had periods as short as 7
seconds in the $1.3~M_\odot$ model and as long as 80 seconds for the
$0.6~M_\odot$ model. Unlike the PNN model, only lower-order modes were
excited. The $n=-1$ and $n=-2$ modes are excited at some point in every model,
while the $n=-3$ mode is excited in the $1.3~M_\odot$ and enriched
$1.0~M_\odot$ models only. In the $1.0~M_\odot$ and $1.3~M_\odot$ models, only
the $n=-1$ mode exhibits short enough growth timescales for the mode to grow
by several $e$-foldings before it is stabilized, but the $0.6~ M_\odot$ model 
actually excites its $n=-2$ mode earlier and more rapidly than the $n=-1$ mode.

\begin{figure}
  \includegraphics[width=\columnwidth]{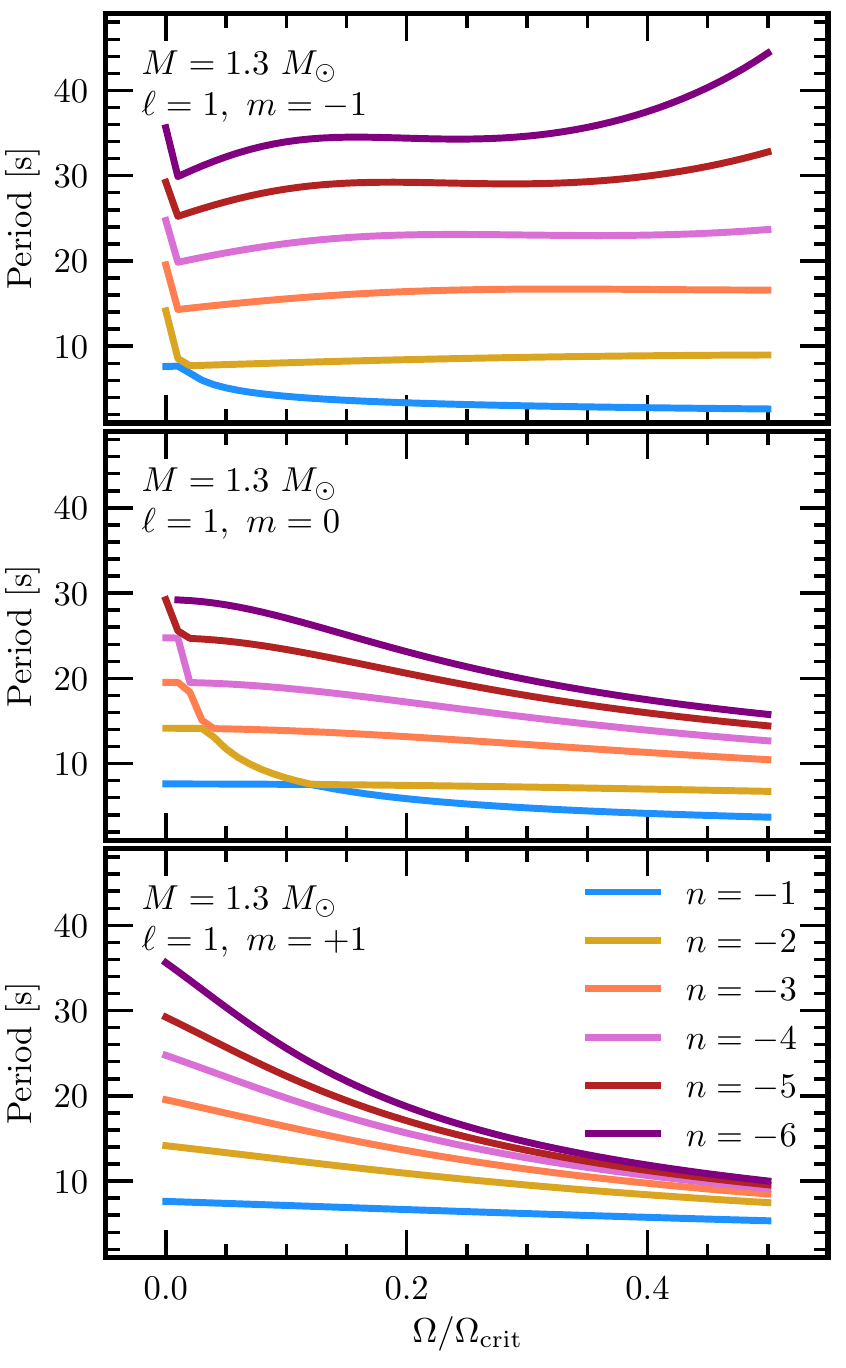}
  \caption{
  Dependence of the \added{inertial (observer's frame)} mode periods on the rotation rate as a fraction of the
  critical rotation rate for the $m=-1$ (top), $m=0$ (middle), and $m=+1$
  (bottom) modes in our $1.3~M_\odot$ fiducial model. At any given rotation 
  rate, the mode with a period of 8--9 seconds is excited with a growth
  timescale of around 2.5 days. For the $m=+1$ case, this is always the $n=-1$
  mode, but due to avoided crossings, the excited mode switches to the $n=-2$ 
  mode in the $m=-1$ and $m=0$ cases at a modest fraction of the critical 
  rotation frequency. No other modes are ever excited on timescales shorter
  than or comparable to the SSS lifetime.}  
  \label{fig:rot1}
\end{figure}

The general trend is that more massive WDs exhibit shorter periods and
shorter growth times. We find that metal enrichment has little effect on the
mode periods, but it significantly reduces growth timescales and the duration
of the SSS phase.

The models made in \mesa{} \texttt{star} are non-rotating, but we can probe
the effects of rotation on the mode periods and growth timescales by using the
traditional approximation \citep{1996ApJ...460..827B, 2005MNRAS.360..465T}.
\added{Note that we do \emph{not} assume Cowling's  approximation (neglecting
the Eulerian perturbation of the gravitational  potential) in rotating or 
non-rotating analyses. Typically Cowling's approximation is assumed along with the
traditional approximation, but in this case it makes little difference since
the Coriolis force only appreciably affects high-order, long-period modes, 
whose frequencies are not greatly affected by the Cowling approximation.}

We investigated how the periods and growth times for $\ell=1$ modes changed in
response to varying the rotation rate $\Omega$ in our fiducial 1.3 $M_\odot$
model. Figure \ref{fig:rot1} shows how periods of $\ell=1$ modes are affected
by rotation up to an $\Omega$ of half of the critical rotation rate 
\replaced{$\Omega_{\mathrm{crit}} = \sqrt{8 G M / (27 R^3)}$}{
$\Omega_{\mathrm{crit}} = \sqrt{8 G M / (27 R^3)}\approx 1\,\textrm{Hz}$}. We now summarize 
the results.

Higher-order zonal ($m=0$) and prograde ($m=1$) modes' periods decreased
modestly with increasing $\Omega$, but for higher-order retrograde ($m=-1$)
modes, periods increased modestly after an initial drop due to a series of
avoided crossings. However, across all $\Omega$'s, there was only ever one mode
excited on timescales comparable to or shorter than the nova evolution
timescale. The period of this mode is 8--9 seconds and its growth timescale is
2.5 days, in great agreement with the non-rotating results shown in
Figure~\ref{fig:taus1}. Due to the avoided crossings, this mode changes in
radial order from $n=-1$ to $n=-2$ at about 2\% and 12\% of
$\Omega_{\mathrm{crit}}$ for the $m=-1$ and $m=0$ cases, respectively. With no
significant change in the periods of the excited mode, we expect no observable
effect from rotation on these oscillations other than incidental effects
rotation may have on the accretion and runaway processes.


%% file: discussion.tex
\section{Comparison to Observation} 
\label{sec:discussion}
The goal of this work was to explain the oscillations in post-outburst
novae and persistent supersoft sources described in \citet{2015A&A...578A..39N}
and references therein. We've demonstrated that the $\epsilon$-mechanism is
indeed an effective means to excite $g$-modes with periods similar to those in
observed SSSs. 

However, we have only demonstrated that these modes are unstable in the linear
regime. We cannot predict amplitudes for these oscillations to construct a 
X-ray light curve for comparison. A more complex non-linear calculation
would be required to make such a robust prediction.

Fortunately, our work has confirmed, as expected, that the periods are most
sensitive to the mass of the underlying WD rather than composition or rotation.
Thus, a nova with a known WD mass and observed oscillations would provide a
means to check the efficacy of $g$-modes as a source for these oscillations.
We now review the oscillating post-outburst novae presented in
\citet{2015A&A...578A..39N} and compare them to our models.

\subsection{RS Ophiuchi} 
\label{sub:rs_ophiuchi}
RS Ophiuchi (RS Oph) is a recurrent nova with recurrence times as short as
nine years. From spectral measurements, \citet{2009A&A...497..815B} find a best
orbital solution for a WD with a mass in the range of $1.2-1.4~M_\odot$. From
the recurrence time alone, models from \citet{2013ApJ...777..136W} limit the
WD mass to $M>1.1~M_\odot$, while the effective temperature and duration of the supersoft phase are
most consistent with models with a mass near $1.3~M_\odot$.

However, according to \citet{2015A&A...578A..39N}, RS Oph has oscillations with
a period of 35 seconds, which is significantly longer than the $\approx 6-10$
second periods seen in the $n=-1$ mode of our $1.3~M_\odot$ model. Even giving
a generously low mass of $1.0~M_\odot$ would require exciting the mode only at
late times when it is already stabilizing or by tapping into the $n=-2$
mode during the brief duration that it is unstable.

\subsection{KT Eridani} 
\label{sub:kt_eridani}
KT Eridani (KT Eri) is a nova that also exhibited oscillations with
periods of roughly 35 s at multiple times in its supersoft evolution
\citep{2010ATel.2423....1B, 2015A&A...578A..39N}. \citet{2012A&A...537A..34J}
estimate from the supersoft turn-on time and possible presence of neon
enrichment, the mass of the underlying white dwarf is $1.1~M_\odot \leq
M_{\mathrm{WD}} \leq 1.3~M_\odot$. With a turn-off time of around 300 days
\citep{2011ApJS..197...31S}, models from \citet{2013ApJ...777..136W} are
consistent with this contraint. Similar to RS Oph, the lowest order (and most
easily excited) modes from the 1.0 and 1.3 $M_\odot$ models still cannot 
explain the observed oscillations, but second or third order modes are not out 
of the question if they could be excited.

\subsection{V339 Delphini} 
\label{sub:v339_delphini}

V339 Delphini (V339 Del) is a nova with an observed 54 s oscillation
\citep{2013ATel.5573....1B, 2013ATel.5626....1N}. \citet{2016A&A...590A.123S}
provide an estimate for the ejecta mass of V339 Del of $2-3\times
10^{-5}~M_\odot$. With this and its SSS turn-off time of 150-200 days, V339 Del
is consistent with a WD mass of $M_{\mathrm{WD}}\approx 1.0-1.1~M_\odot$
\citep{2013ApJ...777..136W}. Again returning to our $1.0~M_\odot$ models, we
must rely on even higher order $n < -2$ to explain the observed oscillations.
The $n=-3$ mode is unexcited in the solar composition model, and in the 
metal-enriched model, it is only marginally unstable in that its growth
timescale is comparable to the duration in which it is unstable. Even then the
$n=-3$ mode has a period that is slightly too short during this phase, but
higher-order modes are never excited at all. It is difficult to explain the
oscillations in V339 Del with our models.

\subsection{LMC 2009a} 
\label{sub:lmc_2009a}
LMC 2009a is a recurrent nova, having first been detected in outburst in
1971. From its recurrence time as well as the SSS duration and temperature of
the 2009 outburst, \citet{2016ApJ...818..145B} estimate the mass of the
underlying WD to be $M_{\mathrm{WD}}\approx 1.1~M_\odot-1.3~M_\odot$. The
oscillations during the SSS phase reported in \citet{2014ATel.6147....1N,
2015A&A...578A..39N, 2016ApJ...818..145B} had a period of 33 seconds. With a
similar period and mass estimate to KT Eri, the $g$-mode explanation of these
oscillations is similarly tenuous.

Our models show that metal enrichment does not change mode periods
substantially, and even relatively rapid rotation cannot greatly affect the
periods of excited modes. Rather, it seems that without some more exotic physics
that can couple to higher-order modes, the $g$-modes we see in our model cannot 
adequately explain the oscillations observed in the novae detailed in
\citet{2015A&A...578A..39N}.

\added{The excited modes we did find may indeed be present in the SSS sample.
\citet{2015A&A...578A..39N} only searched for periods in the range 25-100 s, 
which puts the excited modes we found at too short of periods to be detected
given the expected WD masses. Furthermore, the linear analysis we perform
here cannot predict amplitudes, so non-linear affects may damp these
oscillations before they grow strong enough to become observable.}

%% file: conclusions.tex
\section{Conclusions} 
\label{sec:conclusions}
We have used \mesa{} models to confirm the earlier work of 
\citet{1988ApJ...334..220K} on planetary nebula nuclei. We then extended that
work to see what, if any, modes are excited in post-outburst novae via the
$\epsilon$-mechanicsm. In all our models, we found unstable modes with
growth timescales shorter than the lifetime of the post-outburst supersoft
phase.

While metal-enhancement of the WD envelope did expedite the
evolution through the post-outburst phase and the growth of any excited modes,
it did not greatly influence the periods of these modes. Similarly, rotation
only affected the periods of higher-order modes that were not excited, so it
is unlikely to have a strong effect on any oscillations this mechanism might
produce.

Finally, we compared our results to the observed oscillations of several novae.
Broadly, the excited modes we find for comparable nova models have periods that
are too short to explain the observed ocsillations, and neither metal
enhancement nor rotation are sufficient to excite higher-order modes or increase
an excited mode's period.

%% file: appendix_phase.tex
\section{Calculation of Phase Lags} \label{sec:appendix_phase_lags} 
To calculate the sensitivity of the CNO burning rate to density and temperature
perturbations, we followed the method of \citet{1988ApJ...334..220K} with
several changes. For completeness, we outline the entire calculation here.

Thermonuclear burning in the post-outburst nova is dominated by the CNO cycle.
We consider only the basic CN cycle since it produces most of the energy. The
reactions involved are
\begin{eqnarray}
    \pgamma{C}{12}{N}{13} \label{eq:rxn1},\\
    \bdecay{N}{13}{C}{13} \label{eq:rxn2},\\
    \pgamma{C}{13}{N}{14} \label{eq:rxn3},\\
    \pgamma{N}{14}{O}{15} \label{eq:rxn4},\\
    \bdecay{O}{15}{N}{15} \label{eq:rxn5},\\
    \palpha{N}{15}{C}{12} \label{eq:rxn6}.
\end{eqnarray}
We will index the reactants of equations \eqref{eq:rxn1}--\eqref{eq:rxn6} as
1--6. That is, $\iso{C}{12}$ will be denoted by the number 1 in subscripts and
$\iso{N}{15}$ by 6. These indices will be cyclic so that $1-1=6$ and $6+1=1$.

For an isotope $i$ that is both produced and destroyed via proton captures, the
total number of ions of isotope $i$ is represented by $N_i$. Then the net rate
of production of these isotopes is
\begin{equation}
	\label{eq:rate}
    \frac{DN_i}{Dt} = - N_i n_p \sigv_i + N_{i-1} n_p \sigv_{i-1},
\end{equation}
where $D/Dt$ is the Lagrangian time derivative, $n_p$ is the number density of protons, and the $\sigv$'s are the
thermally-averaged reaction rates. If the isotope is created via a beta decay,
the second term is replaced by $N_{i-1}\lambda_{i-1}$ where $\lambda_{i-1}$ is
the decay rate of isotope $i-1$. Similarly,
if the isotope is destroyed by a beta decay, then we replace the first term in
\eqref{eq:rate} with $-N_i \lambda_i$. The total number of ions of isotopes is
related to its mass fraction $X_i$ and mass number $A_i$ via $N_i \propto 
X_i/A_i$. Thus we can rewrite \eqref{eq:rate} in terms of the mass fraction via
\begin{equation}
	\label{eq:rate_lag}
  \frac{DN_i}{Dt} \propto \frac{1}{A_i} \frac{DX_i}{dt}.
\end{equation}
For simplicity, we also introduce a generalized destruction rate, $K_i$ that is
$\lambda_i$ for isotopes destroyed via beta decay and $n_p\sigv_i$ for those
destroyed by proton captures. This gives a generalized rate equation of

\begin{equation}
    \label{eq:rate_gen}
    \frac{DX_i}{Dt} = -X_i K_i + \frac{A_i}{A_{i-1}} X_{i-1} K_{i-1}.
\end{equation}
In the background equilibrium state, these rates all vanish once the mass
fractions have settled to the preferred configuration. Now we introduce
Lagrangian perturbations (denoted by the $\delta$ symbol) in temperature and
density with frequency $\sigma$, \explain{All subsequent $\sigma$'s have their signs flipped relative to the first submission to reflect proper conventions.}
\begin{equation}
    \label{eq:pert}
    \rho \to \rho_0 + \delta \rho\,e^{-i\sigma t} \qquad T \to T_0 + \delta T\,
    e^{-i\sigma t},
\end{equation}
where subscripts of 0 indicate the constant equilibrium values. The generalized destruction rates, $K_i$ will also change, but only for 
reactions involving proton captures:
\begin{equation}
    \label{eq:pert_K}
    K_i = \lambda_i \to \lambda_{i,0} \qquad K_i = n_p\sigv_i \to K_{i,0} +
    K_{i,0}\left[\frac{\delta \rho}{\rho} + \nu_i\frac{\delta T}{T}\right]
    e^{-i\sigma t},
\end{equation}
where $\nu_i = d\ln\sigv_i/d\ln T$. Similarly, the
mass fractions $X_i$ and their derivatives will also change:
\begin{equation}
    \label{eq:pert_frac}
    X_i \to X_{i,0} + \delta X_{i}\,e^{-i\sigma t} \qquad \frac{DX_i}{Dt} \to 
    -i \sigma \delta X_i\,e^{-i\sigma t}.
\end{equation}
Phase lags will only be present if the values of $\delta X_i$ are complex. Now
applying the perturbations of \eqref{eq:pert_K} and \eqref{eq:pert_frac} to 
\eqref{eq:rate_gen}, subtracting off the equilibrium solution, and dividing out
the exponential dependence gives 
\begin{equation}
  \label{eq:rate_pert} -i\sigma\delta X_i = -\left(\delta X_{i} K_{i, 0}
  + X_{i,0}\delta K_{i}\right) + \frac{A_i}{A_{i-1}}\left(\delta X_{i-1} 
  K_{i-1, 0} + X_{i-1,0}\delta K_{i-1}\right),
\end{equation}
where we've left the perturbation of the generalized rate as a generic 
$\delta K_i$. Specializing to the three classes of isotopes (creation by 
beta decay, destruction by beta decay, or no beta decays) and noting that by
conservation of mass,
\begin{equation}
  \label{eq:conservation} \frac{A_i}{A_{i-1}}
  \left(\delta X_{i-1} K_{i-1, 0} + X_{i-1,0}\delta K_{i-1}\right) =
  X_{i,0}K_{i,0}\left(\frac{\delta X_{i-1}}{X_{i-1,0}} + \frac{\delta K_{i-1}}
  {K_{i-1,0}}\right),
\end{equation}
we get
\begin{eqnarray}
  i\frac{\sigma}{K_{i,0}} \frac{\delta X_i}{X_{i,0}} &=& 
  \left(\frac{\delta X_i}{X_{i,0}} + 
  \frac{\delta K_i}{K_{i,0}}\right) - \left(\frac{\delta X_{i-1}}{X_{i-1,0}} + 
  \frac{\delta K_{i-1}} {K_{i-1,0}}\right),\\
  \label{eq:pert_gen} \frac{ K_{i,0} - i\sigma}{K_{i,0}}
  \frac{\delta X_i}{X_{i,0}} - \frac{\delta X_{i-1}}{X_{i-1,0}} &=&
  \frac{\delta K_{i-1}}{K_{i-1,0}} - \frac{\delta K_i}{K_{i,0}},\\
  \label{eq:pert_papa} \frac{K_{i,0} - i\sigma}{K_{i,0}} 
  \frac{\delta X_i}{X_{i,0}} - \frac{\delta X_{i-1}}{X_{i-1,0}} &=&
  (\nu_{i-1} - \nu_i) \frac{\delta T}{T_0} \qquad (i = 1,4),\\
  \label{eq:pert_pab} \frac{\lambda_i - i\sigma}{\lambda_{i}}
  \frac{\delta X_i}{X_{i,0}} - \frac{\delta X_{i-1}}{X_{i-1,0}} &=& 
  \frac{\delta \rho}{\rho_0} + \nu_{i-1}\frac{\delta T}{T_0} \qquad (i = 2,5),\\
  \label{eq:pert_bpa} \frac{K_{i,0} - i\sigma}{K_{i,0}} 
  \frac{\delta X_i}{X_{i,0}} - \frac{\delta X_{i-1}}{X_{i-1,0}}&=&
  - \frac{\delta \rho}{\rho_0} - \nu_i\frac{\delta T}{T_0} \qquad (i = 3,6).
\end{eqnarray}
Here \eqref{eq:pert_gen} is still a general result while \eqref{eq:pert_papa} - 
\eqref{eq:pert_bpa} relate the relative mass fraction perturbations to the 
equilibrium conditions and the temperature and density perturbations for
isotopes that are created and destroyed by proton captures
\eqref{eq:pert_papa}, created by proton captures and destroyed by beta decays
\eqref{eq:pert_pab}, and created by beta decays and destroyed by proton
captures \eqref{eq:pert_bpa}. These constitute a set of six equations in six
unknowns. For a given temperature, density, and equilibrium set of abundances,
we can then query the \texttt{rates} module of \mesa{} to get $\lambda_i$,
$K_{i,0}(\rho_0, T_0)$, and $\nu_i(T_0)$ to get an expression for $\delta X_i$
in terms of $\sigma$, $\delta T/T_0$, and $\delta \rho/\rho_0$. In general,
this has the form

\begin{equation}
  \label{eq:alphabeta}\frac{\delta X_i}{X_{i,0}} = 
  \left(\alpha\frac{\delta \rho}{\rho_0} + \beta\frac{\delta T}{T_0}\right)
  e^{-i\sigma t},
\end{equation}
where the $\alpha$'s and $\beta$'s come from solving the system of equations
above. They depend only on the various $K_i$'s, $\nu_i$'s, and $\sigma$. They
are in general complex, giving rise to phase delays between the
temperature/density perturbation and the actual changes in abundances.
\citet{1988ApJ...334..220K} solved for these $\alpha$'s and $\beta$'s explicitly
in the limit where beta decays occur much more quickly than proton captures.
This limit is valid in the case of a PNN, but at the higher temperatures present
in some of the post-outburst novae, this assumption fails, so the full matrix
inversion calculation is needed to solve for these quantities.

To see
how this affects wave excitation via the $\epsilon$-mechanism, we need to relate
these $\alpha$'s and $\beta$'s to the nuclear energy generation rate.
The energy generation rate due to the destruction of species $i$ is given by
\begin{equation}
  \label{eq:eps_i} \epsilon_i = \frac{X_i K_i Q_i}{A_i m_p},
\end{equation}
where $K_i$ is again the generalized destruction rate and $Q_i$ is the energy
released by the destruction of one isotope (roughly the difference in binding
energies). Then the total energy generation rate is just the sum over all of
these rates. After accounting for the perturbations in $K_i$ and $X_i$, the
perturbation in the overall energy generation rate is
\begin{equation}
  \label{eq:eps_tot} \frac{\delta \epsilon}{\epsilon_0} = \left(A 
  \frac{\delta \rho}{\rho_0} + B\frac{\delta T}{T_0}\right)e^{-i\sigma t},
\end{equation}
where
\begin{equation}
  \label{eq:A} A = \frac{d\ln \epsilon}{d\ln \rho} = 
  \frac{\left(\sum_i\alpha_i\epsilon_i\right) + \epsilon_1 + \epsilon_3 +
  \epsilon_4 + \epsilon_6}{\epsilon_0}
\end{equation}
and
\begin{equation}
  \label{eq:B} B = \frac{d\ln \epsilon}{d\ln T} = 
  \frac{\left(\sum_i\beta_i\epsilon_i\right) + \nu_1\epsilon_1 + 
  \nu_3\epsilon_3 + \nu_4\epsilon_4 + \nu_6\epsilon_6}{\epsilon_0}.
\end{equation}

In the long-period limit $\sigma\to0$, we expect $A\to 1$, but in general, $A<1$
for periods in the 1-1000 second range. Similarly, $B$ is smaller than the
expected unperturbed value for periods in this range, causing an enhanced
stability in the burning rate with changes to temperature and densities.

\begin{figure}[h]
  \plottwo{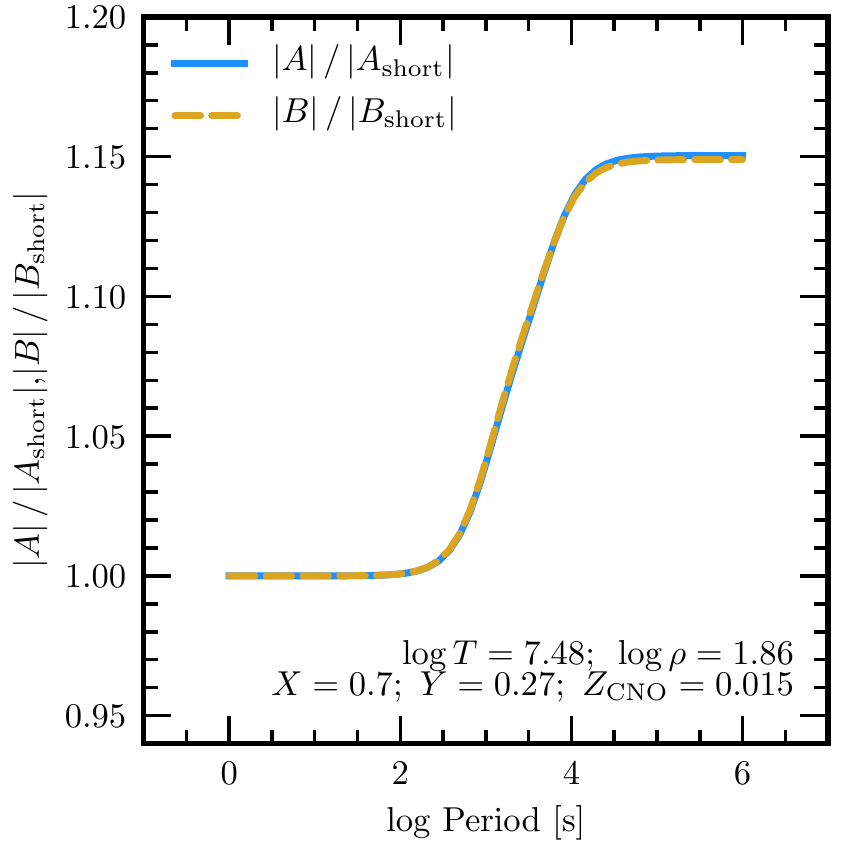}{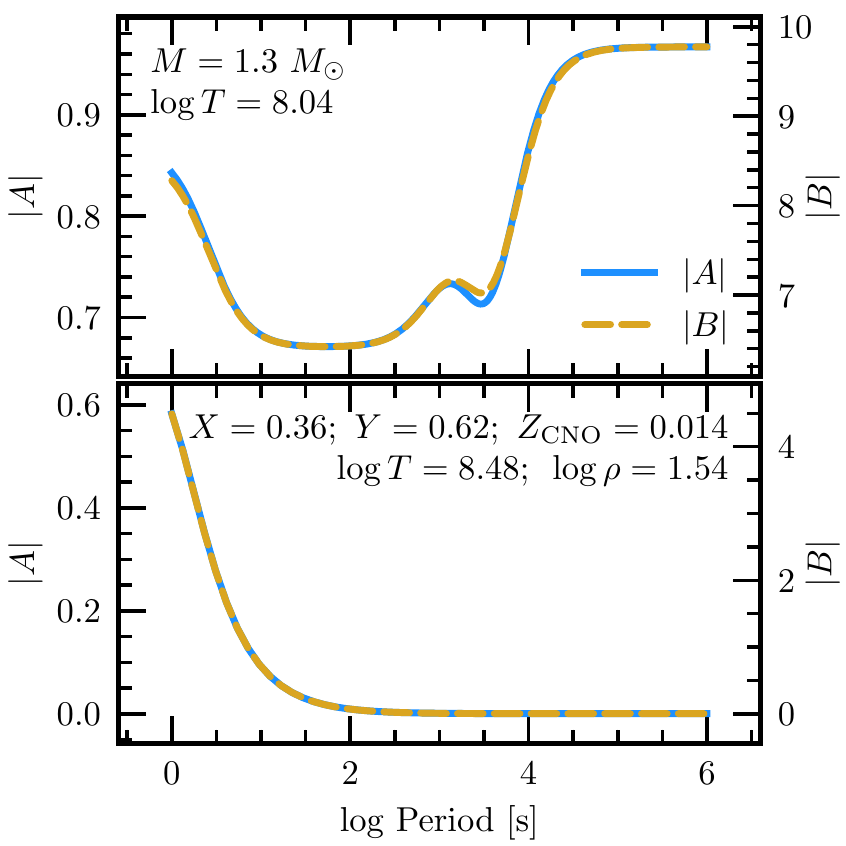}

  \caption{\textbf{Left:} dependence of the logarithmic derivatives of the 
  nuclear energy generation rate (relative to the \textbf{short}-period limit) on
  pulsation period. This figure corroborates the similar Figure 3 from
  \cite{1988ApJ...334..220K}. \textbf{Right:} the actual values of these
  derivatives for our fiducial $1.3~M_\odot$ post-outburst nova model (top) 
  and for a much hotter ($3\times 10^8$ K) model with the same composition
  and density.}
  
  \label{fig:ab}
\end{figure}

As a simple
check that our method is consistent with the work of
\cite{1988ApJ...334..220K}, the left panel of Figure \ref{fig:ab} reproduces Figure 3 of that
paper, where the derivatives $A=d\ln\epsilon_{\mathrm{nuc}}/d\ln\rho$ and
$B=d\ln\epsilon_{\mathrm{nuc}}/d\ln T$ are plotted for a particular
temperature, density, and composition. We find excellent agreement with our
general approach. The right panel of Figure~\ref{fig:ab} shows how the actual
values of $A$ and $B$ vary as a function of period for our fiducial 
$1.3~M_\odot$ post-outburst nova model as well as a much hotter version of that
model, demonstrating that temperature and density sensitivity indeed vanish
at such high temperatures as the reaction cycle becomes limited by beta decays.

Generally, $A$ and $B$ are local quantities since they depend on the local
equilibrium values for the $X_i$, $\rho$ and $T$. Since we needed values for 
$A$ and $B$ at a large range of periods for computations with \gyre{} and for
every snapshot saved during the post-outburst phase, we decided to simply sample
the point of peak CNO burning and apply the modified values of $A$ and $B$ to
all regions with significant burning. The area of peak burning is what drives 
the $\epsilon$-mechanism, so this is the value and location that matters
most.

To incorporate the phase lags defined above, we modify \gyre{} so that the
$\epsilon_{\rm ad}$ and $\epsilon_{S}$ partial derivatives are evaluated via the 
expressions

\begin{align}
\epsilon_{\rm ad} &\equiv \left( \frac{\partial\ln\epsilon}{\partial\ln P} \right)_{S} =  \frac{A}{\Gamma_{1}} + \nabla_{\rm ad} B,\\
\epsilon_{\rm S} &\equiv c_{P} \left( \frac{\partial\ln\epsilon}{\partial S} \right)_{P} = - \upsilon_{T} A + B,
\end{align}
(all symbols have the same meaning as in \citet{Townsend2017}). For efficiency reasons, the complex coefficients $A$ and $B$ are 
pre-calculated on tables spanning a range of periods, and interpolated at 
runtime using cubic splines. These new capabilities will be included in version
5.1 of \gyre{}.